\documentclass{aastex63}
\usepackage{geometry}               
\usepackage{graphicx}
\usepackage{amssymb}
\usepackage{epstopdf}
\usepackage{fullpage}
\usepackage{natbib}
\usepackage{subfigure}
\usepackage{epsfig}
\usepackage{multirow}
\usepackage[figuresright]{rotating}
\usepackage{amsmath}
\usepackage{url}

\def\aj{AJ}                   
\def\apj{ApJ}                 
\def\apjl{ApJL}                
\def\apjs{ApJS}               
\def\ao{Appl.~Opt.}           
\def\aap{A\&A}                

\def\icarus{Icarus}           

\def\jgr{J.~Geophys.~Res.}    
\def\jqsrt{J.~Quant.~Spec.~Radiat.~Transf.}

\def\planss{Planet Space Sci.}   

\def\james{J. Adv. Model Earth Sy.}     

\shorttitle{Energy Budgets for Exoplanets}
\shortauthors{Shields et al.}

\begin{document}
\title{Energy Budgets for Terrestrial Extrasolar Planets}
\correspondingauthor{Aomawa Shields}
\email{shields@uci.edu}

\author{Aomawa L. Shields}
\affil{Department of Physics and Astronomy \\
University of California, Irvine \\
4129 Frederick Reines Hall \\
Irvine, CA\\ 
92697-4575 USA}

\author{Cecilia M. Bitz}
\affiliation{Department of Atmospheric Sciences\\
University of Washington\\
Seattle, WA\\
98105-6698 USA\\
\\
\textit{Submitted 2019 August 16; accepted 2019 September 14}\\}

\author{Igor Palubski}
\affiliation{Department of Physics and Astronomy \\
University of California, Irvine \\
4129 Frederick Reines Hall \\
Irvine, CA\\ 
92697-4575 USA}



%

\begin{abstract}
The pathways through which incoming energy is distributed between the surface and atmosphere has been analyzed for the Earth. However, the effect of the spectral energy distribution of a host star on the energy budget of an orbiting planet may be significant given the wavelength-dependent absorption properties of atmospheric CO$_2$ and water vapor, and surface ice and snow. We have quantified the flow of energy on aqua planets orbiting M-, G-, and F-dwarf stars, using a 3D Global Climate Model with a static ocean. The atmosphere and surface of an M-dwarf planet receiving an instellation equal to 88\% of the modern solar constant at the top of the atmosphere absorb 12\% more incoming stellar radiation than those of a G-dwarf planet receiving 100\% of the modern solar constant, and 17\% more radiation than a F-dwarf planet receiving 108\% of the modern solar constant, resulting in climates similar to modern-day Earth on all three planets, assuming a 24-hr rotation period and fixed CO$_2$. At 100\% instellation, a synchronously-rotating M-dwarf planet exhibits smaller flux absorption in the atmosphere and on the surface of the dayside, and a dayside mean surface temperature that is 37 K colder than its rapidly-rotating counterpart. Energy budget diagrams are included to illustrate the variations in global energy budgets as a function of host star spectral class, and can contribute to habitability assessments of planets as they are discovered. \end{abstract}

\section{Introduction} \label{sec:intro}
A planet's climate is largely determined by its global energy balance between the incoming stellar (shortwave) radiation and the outgoing thermal (longwave) radiation emitted to space. This delicate balance is achieved through a combination of reflected, absorbed, and/or emitted shortwave and longwave radiation throughout a planet's atmosphere and surface. The global energy budget of the Earth has been calculated, based on both numerical simulations and direct observations, and its partitioning is often depicted graphically in what is known as a ``Trenberth diagram" \citep{Kiehl1997, Fasullo2008a, Trenberth2009, Stephens2012}. Analyses of Earth's energy budget has led to a deeper understanding of the various pathways through which incoming energy from a host star flows to and is cycled by a planet's atmosphere and surface, and has been used to identify sources of imbalance in the Earth system, such as that which currently exists as a result of anthropogenic CO$_2$ emissions \citep{Hansen2005, Hansen2011, Trenberth2014}. Trenberth diagrams have also been calculated for Venus, Mars, Jupiter, Titan, and the ``hot Jupiter" exoplanet HD 189733b \citep{Read2016}. 

The interaction between the spectral energy distribution (SED) of a host star and the atmospheres and surfaces of orbiting planets has been shown to strongly affect planetary climate, and these effects depend on the amount of flux emitted by stars in certain wavelength regions \citep{Shields2013, vonParis2013, Shields2014, Godolt2015, Wolf2017a, Shields2018}. Given the wavelength-dependent absorption properties of atmospheric gases as well as surface types, the interaction between host star SED and a planet's atmosphere and surface will therefore influence the various pathways and associated energy budgets of orbiting planets, and the manner in which global energy balance is attained. The atmospheres and icy surfaces of planets orbiting M-dwarf stars have been found to absorb more radiation overall than those of their counterparts orbiting stars with more visible and UV output at equivalent stellar flux distances and with equal rotation periods, resulting in warmer surface temperatures and greater climate stability on M-dwarf planets \citep{Shields2013, Shields2014}.  However, the amount of radiation absorbed versus reflected and other components of the energy budget on an orbiting planet as a function of its host star's SED has not been quantified, yet the energy budget controls how the surface temperature and climate will evolve over time for planets in different stellar environments. 

In this work we have simulated the climates of planets orbiting M-, G-, and F-dwarf stars, and calculated energy budgets that track the flow of shortwave and longwave radiation throughout each planet. We provide these energy budgets in the form of schematic diagrams. We identified the amount of stellar radiation incoming at the top of the atmosphere (hereafter ``instellation") necessary to yield global mean surface temperatures approximating that of modern-day Earth as a function of host star spectral type, and compared the energy budgets for these planets to isolate the effect of stellar SED on their pathways to similar climates.

Planets orbiting close in to their stars will experience strong tidal effects, potentially resulting in captures into spin-orbit resonances \citep{Dole1964} such as 1:1 spin-orbit resonance, where the substellar point of the planet is fixed with time (synchronous rotation). Such a state will significantly impact a planet's climate \citep{Joshi1997, Merlis2010, Edson2011, Showman2011a, Showman2011b, Heng2012, Showman2013, Yang2014, Kaspi2015}. We also simulated the climate of a synchronously-rotating M-dwarf planet receiving 100\% of the modern solar constant to examine the effects of this extreme rotational state on global energy budgets and pathways compared with its rapidly-rotating counterpart.  

In Section 2 we describe the model and methods used to simulate the climates of planets orbiting stars of different spectral types. We present results from these simulations, including Trenberth diagrams and a table of energy budgets, in Section 3. An analysis and discussion of the differences in energy budgets for the planets explored is provided in Section 4. Conclusions follow in Section 5.

\section{Methods and Models} \label{sec:models}
We used version 4 of the Community Climate System Model (CCSM4), a three-dimensional (3D) global climate model (GCM) developed to simulate and predict climate and weather patterns on the Earth \citep{Gent2011}. CCSM4 contains an atmospheric component (The Community Atmosphere Model version 4, or CAM4) and the Los Alamos sea ice model (CICE version 4; \citeyear{Hunke2008}). We ran simulations with a 50-meter deep slab ocean without heat flux, but treated as fully mixed with depth. This suite of coupled model components has been used in previous work (see e.g., \citealp{Bitz2012, Shields2013, Shields2014, Shields2016a, Shields2018}). The horizontal angular resolution is nominally 2$^\circ$. We modified the percentages of incoming stellar flux in each of the twelve wavelength bands that are input to CAM4 according to the SEDs of G2V star The Sun  \citep{Chance2010}, M3V star AD Leo\footnote{\url{http://vpl.astro.washington.edu/spectra/stellar/mstar.htm}} \citep{Reid1995, Segura2005}, and F2V star HD128167\footnote{\url{http://vpl.astro.washington.edu/spectra/stellar/other_stars.htm}} \citep{Segura2003}. For full details on how the model has been applied to exoplanets, see Shields \emph{et al.} (\citeyear{Shields2013}). 

We simulated the climates of M-, G-, and F-dwarf aqua planets (no land) receiving a range of instellations from their host stars, assuming circular orbits, a radius, mass, and obliquity equal to the Earth's, and atmospheres with 1-bar surface pressure and Earth-like levels of CO$_2$. Water vapor was permitted to adjust during each simulation in accordance with standard evaporation and precipitation processes on the surface and in the atmosphere. We simulated Earth-like (24-hr) rotation periods and also a synchronous rotation period (obliquity = 0) for the M-dwarf planet. We identified the level of instellation required from M- and F-dwarf host stars to generate climates similar to modern-day Earth.   

As done in Shields \emph{et al.} (\citeyear{Shields2013, Shields2014, Shields2016a, Shields2018}), we used the sea-ice albedo parameterization of CCSM3, as it is easier to manipulate than later versions. This parameterization divides the surface albedo into two bands, visible ($\lambda \leqslant$ 0.7 $\mu$m) and near-IR ($\lambda >$ 0.7 $\mu$m). The default near-IR and visible band albedos, tuned for a solar spectrum, are 0.30 and 0.67 for cold bare ice, and 0.68 and 0.80 for cold dry snow, respectively. For our simulations of M- and F-dwarf planet climates, we calculated the two-band albedos weighted by the spectrum of each host star. All ice and snow albedos used are provided in Table 1. We present a comparison and analysis of the differences in the planets' global energy budgets in the following sections. 

\linespread{1.0}
\begin{table}[!htp] 
\caption{Two-band albedos employed for ice and snow in the GCM, weighted by the spectrum for G-dwarf star the Sun, M-dwarf star AD Leo, and F-dwarf star HD128167. $E$ and $P$ denote water evaporation and precipitation, respectively.} 
\vspace{2 mm}
\centering \begin{tabular}{c c c} 
\hline\hline 
Host star & $T<0^\circ$C & $E-P<0$ \\  [0.5ex] 
\hline
Band & NIR/VIS & NIR/VIS\\
M-dwarf & 0.18/0.69 & 0.49/0.97 \\ 
G-dwarf & 0.30/0.67 & 0.68/0.80 \\
F-dwarf & 0.27/0.73 & 0.67/0.99 \\[1ex]
\hline 
\end{tabular} 
\label{table:nonlin} 
\end{table}

\section{Results} \label{sec:res}
Figure 1 shows the global mean surface temperature for planets receiving different amounts of instellation from M-, G, and F-dwarf stars, and annual mean energy budget values for planets receiving the amount of instellation to yield equivalent, (modern-day) Earth-like climates. The F-dwarf planet required 108\% of the modern solar constant to generate a global mean surface temperature of 287 K, similar to modern-day Earth and to the G-dwarf planet receiving 100\% of the modern solar constant (288 K). The M-dwarf planet required only 88\% of the modern solar constant to produce a similar climate, with a global mean surface temperature of 287 K (Figure 1a). 

While the F-dwarf planet receives the largest amount of instellation from its host star, its atmosphere also reflects the largest percentage of that incoming instellation\textemdash 2\% more than the G-dwarf planets' atmosphere (Figure 1b) and nearly 12\% more than the M-dwarf planet's atmosphere\textemdash while the M-dwarf planet's atmosphere absorbs the most\textemdash over 15\% more than the G-dwarf's, and nearly 20\% more than the F-dwarf's (Figure 1c). At the surface, the F-dwarf planet also reflects the largest percentage of its incoming radiation, 16\% of the shortwave (SW) that reaches the surface, 5\% and 9\% more than the surfaces of the G- and M-dwarf planets, respectively. In contrast, the M-dwarf planet absorbs the most\textemdash 93\%\textemdash of the SW reaching the surface (Figure 1d), $\sim$4\% more than the G-dwarf planet's surface and 9\% more than the surface of the F-dwarf planet. The M-dwarf planet, whose atmosphere and surface combined absorb 12\% more radiation than the G-dwarf planet and 17\% more radiation than the F-dwarf planet, has a correspondingly larger outgoing longwave radiation (OLR). Energy budget fluxes for all three planets are shown in Table 2. 
\begin{figure}[!htb]
\begin{center}
\includegraphics[scale=0.35]{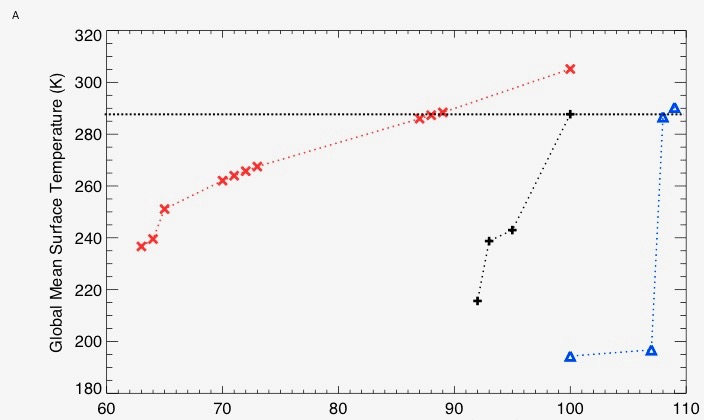}\\
\includegraphics[scale=0.22]{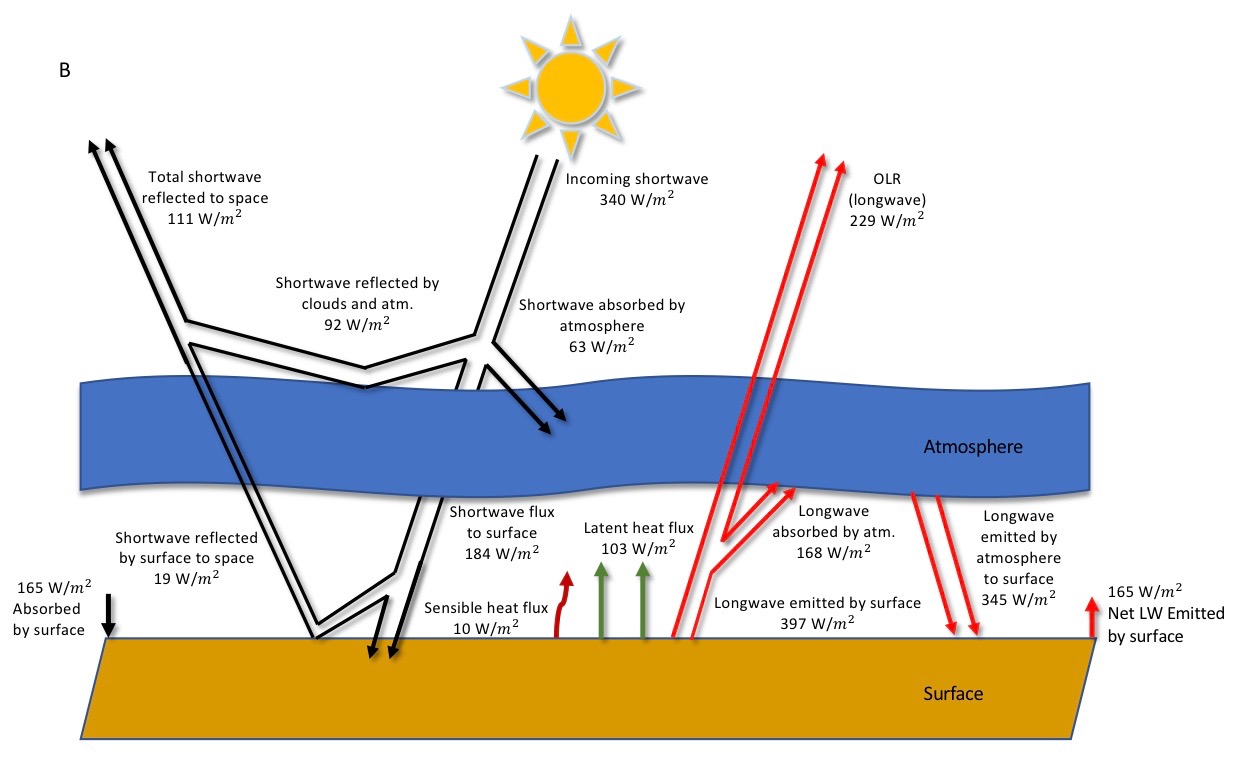}\\
\includegraphics[scale=0.22]{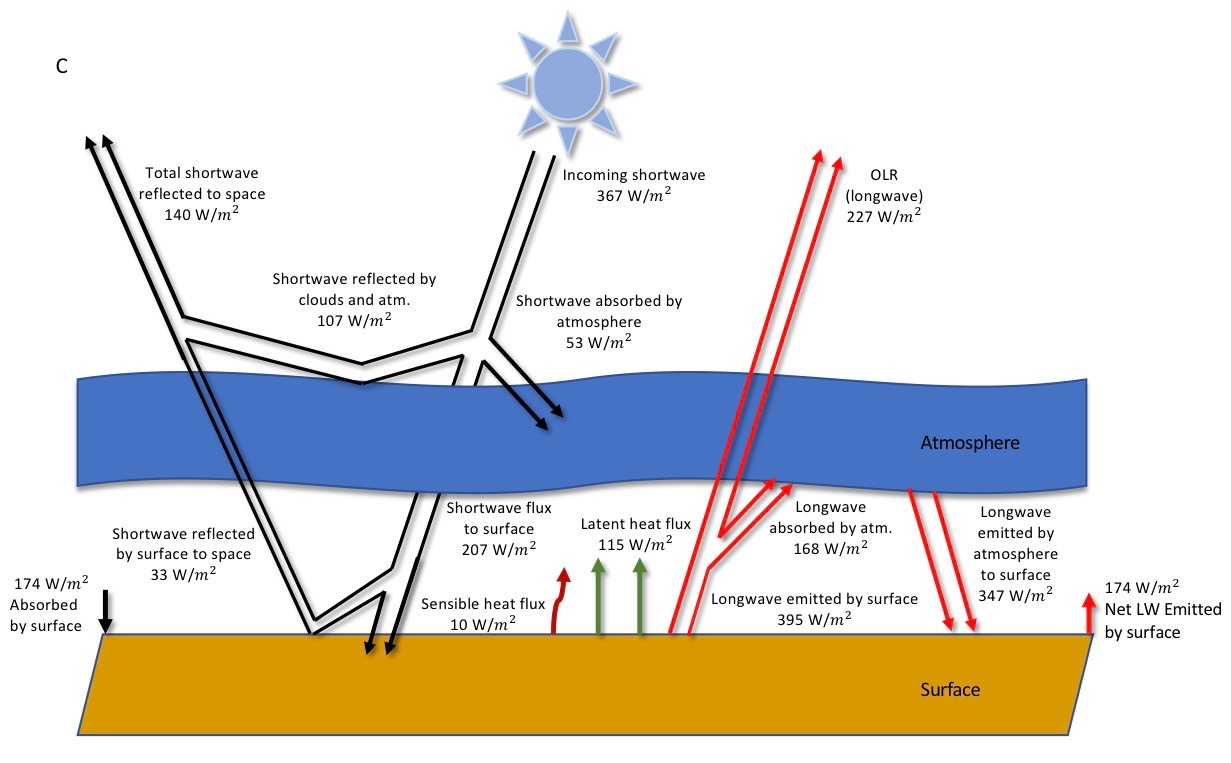}\\
\includegraphics[scale=0.22]{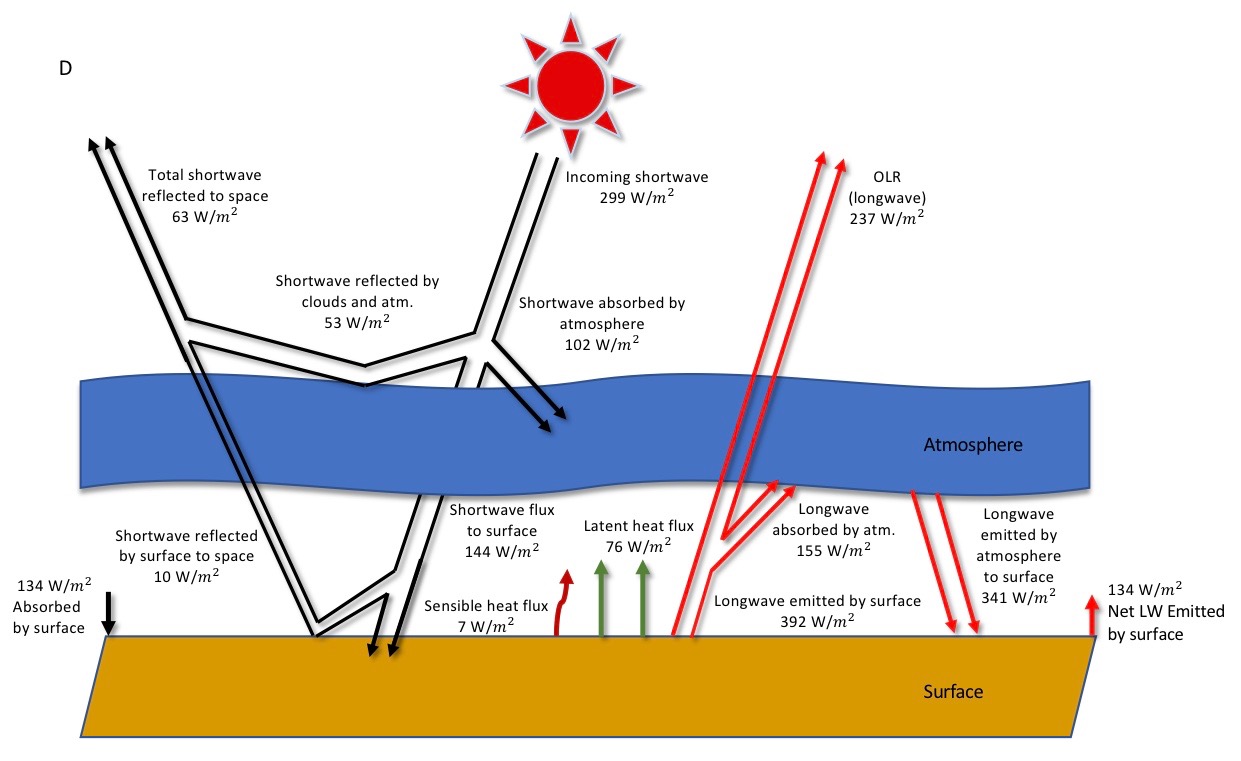}\\
\caption{Top row: Global mean surface temperature as a function of instellation for F- (blue triangles), G- (black plus symbols) and M-(red x's) dwarf terrestrial planets, averaged over 60-120 years of CCSM simulations. The horizontal black dotted line indicates a temperature of $\sim$288 K, similar to modern-day Earth, and aligns closest to the G-dwarf planet at 100\% instellation (second row), the F-dwarf planet receiving 108\% instellation (third row), and the M-dwarf planet at 88\% instellation (fourth row), whose annual mean global energy budgets at the top of the atmosphere and the surface are averaged over 80-100 yrs of CCSM simulations.}
\end{center}
\end{figure}

Figure 2 shows annually-averaged climatic variables across the three planets, all of which have open water in the tropics and mid latitudes. At higher latitudes where there is ice on these planets, the F-dwarf planet has the highest surface albedos (Figure 2a), while the M-dwarf planet has lowest, with a lower ice fraction in these regions (Figure 2b). Surface temperatures are warmer here on the M-dwarf planet (Figure 2c). The higher instellation received by the F-dwarf increases the specific humidity in the tropics and mid latitudes where there is open ocean to absorb strongly (Figure 2d), resulting in slightly warmer temperatures here relative to the other two planets. However, in the high-latitude ice-covered regions the larger absorption of radiation by lower-albedo ice on the M-dwarf planet increases the specific humidity by a larger factor relative to the F-dwarf planet, contributing to significantly increased surface temperatures at higher latitudes and a smaller equator-to-pole temperature contrast on the M-dwarf planet compared to the F- and G-dwarf planets.

Figure 3 shows the annual mean energy budget for an M-dwarf planet with a 24-hr rotation period receiving 100\% of the modern solar constant from its star, the dayside of a synchronously-rotating M-dwarf planet receiving equivalent instellation, and a contour map of the annual mean surface temperature across the synchronous planet. Of the 680 W/m$^2$ received by the dayside of the synchronous planet (1360 W/m$^2$ divided by 2), 26\% is reflected by the atmosphere and 38\% is absorbed, compared with 15\% reflected and 43\% absorbed by that of the M-dwarf planet with a 24-hr rotation period (Figure 3a). While a similar percentage of each planet's incoming SW reaches the surface, 56\% less flux is absorbed by the dayside surface of the synchronously-rotating planet (Figure 3b). However, of the SW absorbed by the entire dayside (atmosphere + surface), only 53\% leaves as OLR. This lower relative thermal emission results in dayside surface temperatures reaching 287 K, and a sizable region above freezing ($>$273 K) at the substellar point. Temperatures do get 66 degrees colder (221 K) on portions of the dayside of the planet (Figure 3c), resulting in ice cover in those annuli. Though this ice present on the dayside has a relatively low albedo, it is still more reflective than ocean, which comprises the entire surface of the rapidly-rotating planet, which never gets below freezing. The dayside mean surface temperature is $\sim$268 K, 37 degrees colder than the rapidly-rotating M-dwarf planet (305 K), which has a narrower  temperature difference (42 K) between maximum (319 K) and minimum (277 K) surface temperatures. The nightside of the planet gets as cold as 218 K, resulting in a global mean surface temperature of 245 K on the synchronously-rotating M-dwarf planet, 60 degrees colder than its rapidly-rotating counterpart.

\linespread{1.0}
\begin{table}[!htp] 
\caption{Selected annual mean radiative fluxes (in W/m$^2$) for planets receiving the instellation necessary from F-, G-, and M-dwarf host stars to yield global mean surface temperatures similar to modern-day Earth. SW is incoming stellar radiative flux. LW is outgoing thermal flux from the planet. An obliquity of 23$^\circ$ and a 24-hr rotation period is assumed for all three planets.} 
\vspace{2 mm}
\centering \begin{tabular}{c c c c c}  
\hline\hline
 & Flux  & F-dwarf & G-dwarf & M-dwarf\\ [0.5ex]
 & Global mean surface temperature (K) & 287 & 288 & 287\\
 & Instellation (percent of the modern solar constant) & 108 & 100 & 88\\
& Incoming & 367 & 340 & 299\\
& Reflected by atmosphere & 107 & 92.2 & 52.7\\
& Percentage of incoming SW  & 29.2\% & 27.1\% & 17.6\%\\ 
SW & Absorbed by atmosphere & 52.9 & 63.3 & 102\\
& Percentage of incoming  SW & 14.4\% & 18.6\% & 34.2\%\\ 
& Reaching surface & 207 & 184 & 144\\
& Reflected (surface) & 33.0 & 19.2 & 9.78\\
& Percentage of SW reaching surface  & 15.9\% & 10.4\% & 6.79\%\\ 
& Absorbed (surface) & 174 & 165 & 134\\
& Percentage of SW reaching surface  & 84.1\% & 89.6\% & 93.2\%\\ [1ex]
\hline
LW & Emitted by surface & 395 & 397 & 392\\
& Absorbed by atmosphere & 168 &168 & 155\\
& Percentage of emitted by surface  & 42.4\% & 42.3\% & 39.6\%\\
& Emitted by atmosphere to surface & 347 & 345 & 341\\
& OLR & 227 & 229 & 237\\
& Percentage of emitted by surface  & 57.5\% & 57.6\% & 60.4\%\\[1ex]
\hline\hline
\end{tabular} 
\label{table:nonlin} 
\end{table}
\pagebreak

\begin{figure}[!htb]
\begin{center}
\includegraphics[scale=0.45]{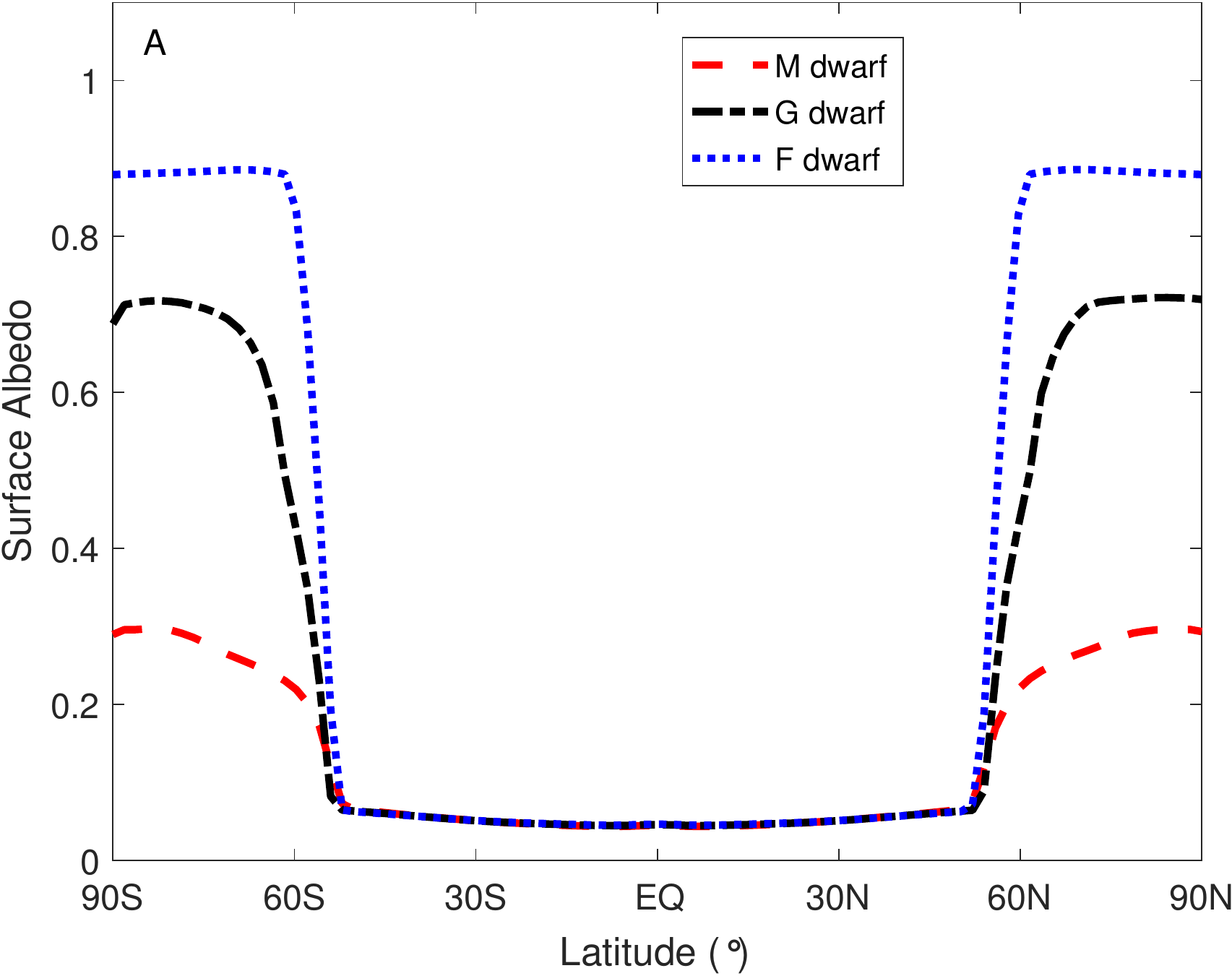}
\includegraphics[scale=0.45]{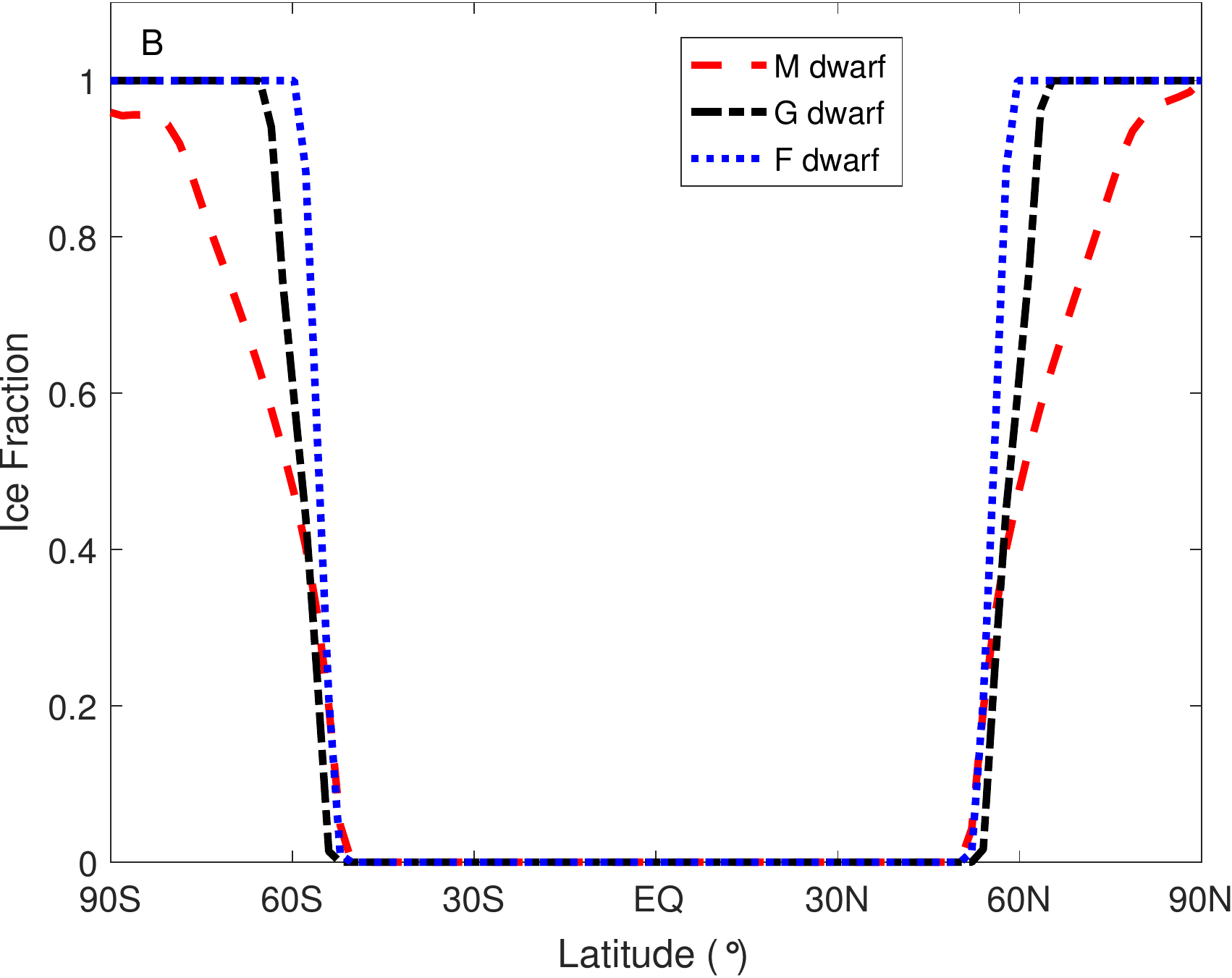}\\
\includegraphics[scale=0.45]{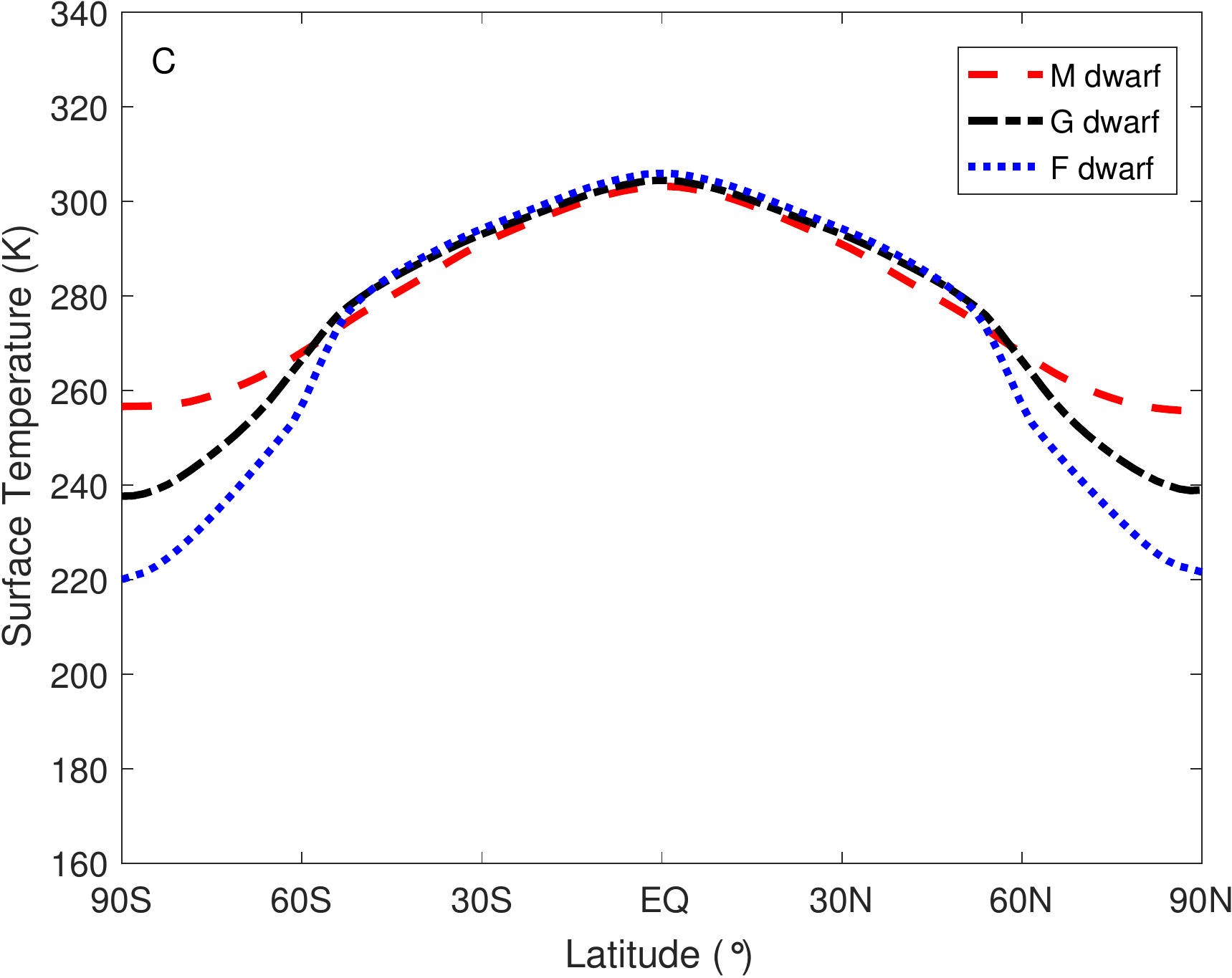}
\includegraphics[scale=0.42]{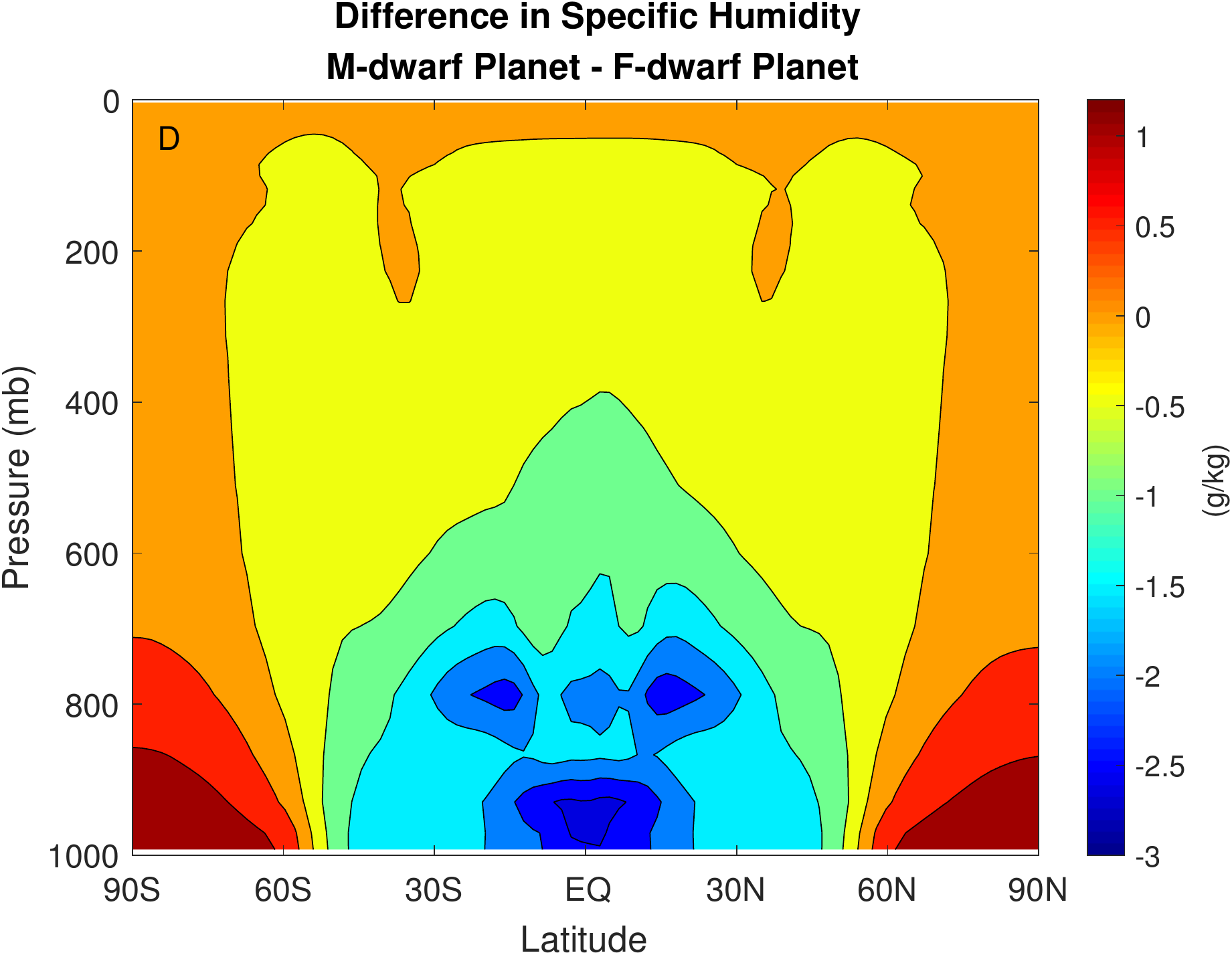}\\
\caption{Annual mean surface albedo, ice fraction, surface temperature, and specific humidity (M-dwarf planet minus F-dwarf planet) for F-, G-, and M-dwarf planets with climates similar to modern-day Earth and receiving 108\%, 100\%, and 88\% of the modern solar constant from their host stars, averaged over 60-100 yrs of CCSM simulations.} 
\label{Figure 2.}
\end{center}
\end{figure}

\begin{figure}[!htb]
\begin{center}
\includegraphics[scale=0.30]{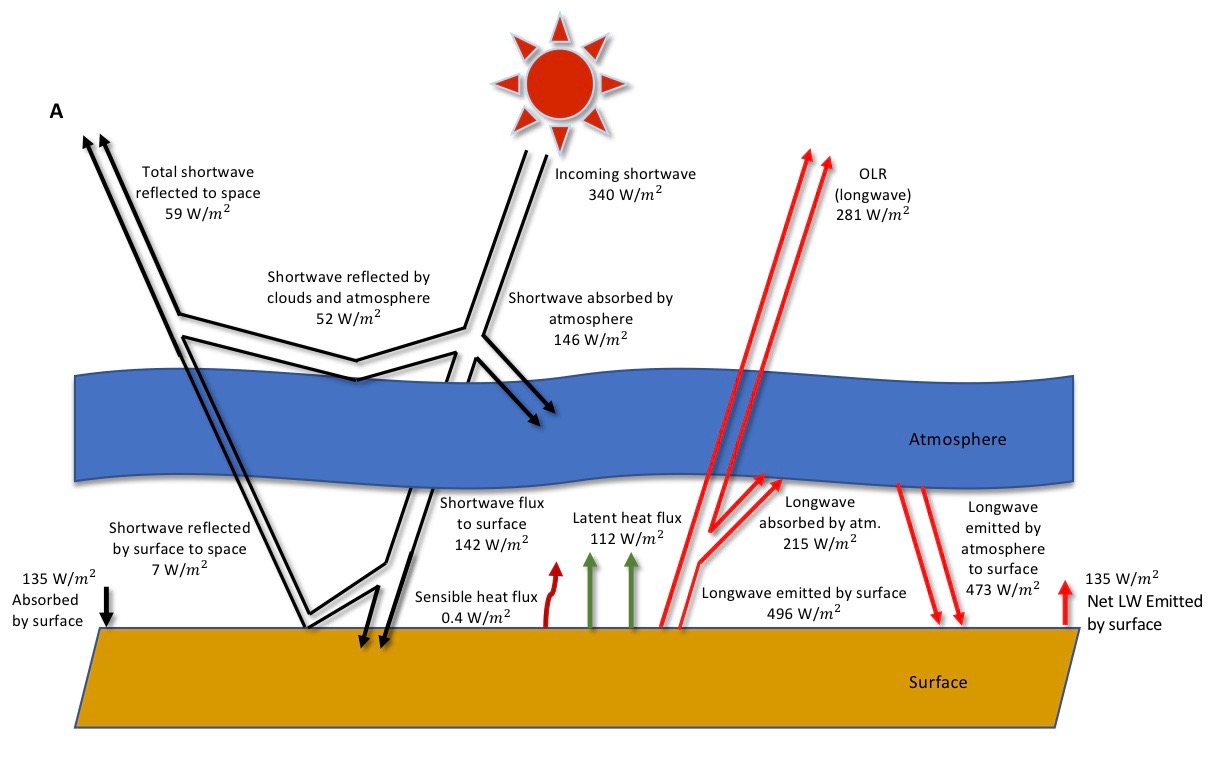}
\includegraphics[scale=0.30]{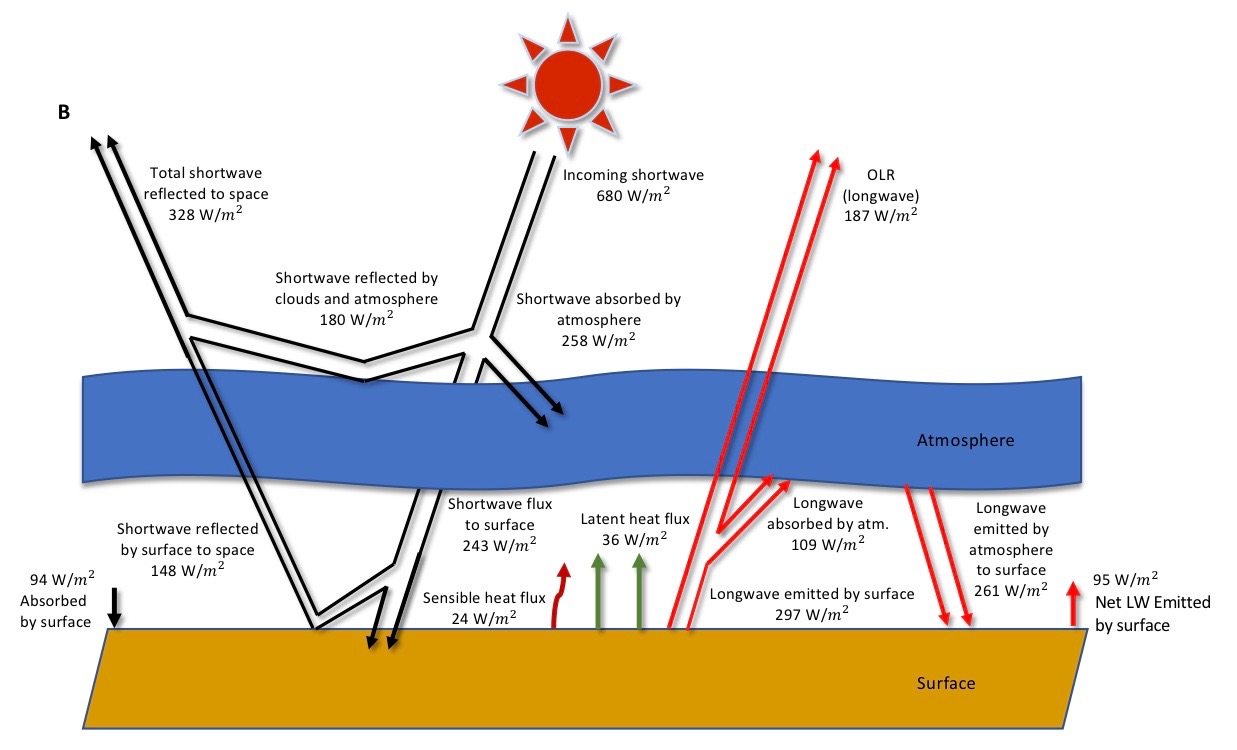}\\
\includegraphics[scale=0.55]{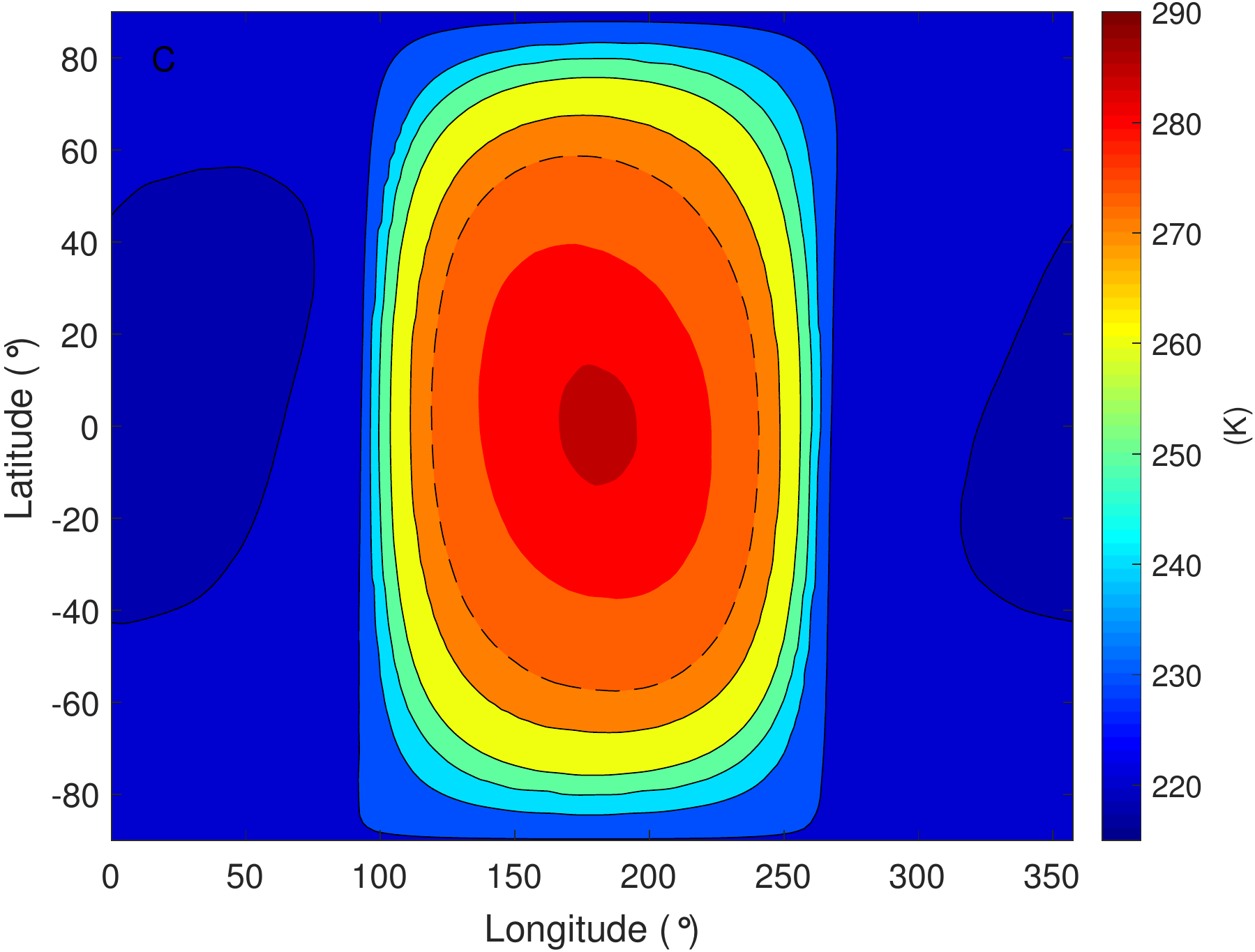}
\caption{Top: Annual mean energy budgets for a terrestrial M-dwarf planet with a 24-hr rotation period receiving 100\% of the modern solar constant from its host star, averaged over 80 yrs of CCSM simulations. Middle: Annual mean energy budgets for a synchronously-rotating M-dwarf planet receiving equivalent instellation. Bottom: Surface temperature as a function of latitude and longitude for the synchronously rotating planet. Zero eccentricity and obliquity are assumed for the synchronous planet. The freezing point (273 K) on the planet is labeled by a dashed contour line. } 
\label{Figure 3.}
\end{center}
\end{figure}

\section{Discussion} \label{sec:disc}

Our results indicate that a host star's spectral type has a significant effect on the energy budget of an orbiting terrestrial planet, assuming fixed CO$_2$ levels and equivalent planetary rotation periods. The interactions between host star SED and gases such as CO$_2$ and H$_2$O in a planet's atmosphere, and icy and snowy surfaces on the ground, all of which have wavelength-dependent absorption properties, are responsible for the difference in energy budgets for planets orbiting different types of stars. Each star-planet interaction produces a unique partitioning of the incoming SW and outgoing LW, resulting in a specific pathway to a planet's climate that varies with host star spectral type. The spectrum dependence of the energy budgets of terrestrial planets, when quantified, reveals that the increased absorption of incoming near-IR radiation by the atmospheres and surfaces of M-dwarf planets causes these planets to be warmer than planets orbiting hotter, brighter stars at equivalent flux distances and require less instellation to produce similar climates, assuming similar planetary rotation periods and atmospheric gas concentrations. 

The amount of instellation required to generate a climate similar to modern-day Earth varies with host star spectral type. The atmosphere and surface of a planet orbiting an F-dwarf star reflects more incoming radiation than those of a G- or M-dwarf planet, requiring 8\% and 20\% more incoming radiation to produce a climate similar to these planets, respectively. The higher albedo of water ice in the visible and near-UV, where the F-dwarf star strongly emits, combined with the reduced atmospheric absorption on this planet, as CO$_2$ and water vapor absorb strongly in the near-IR rather than the visible or near-UV, cause the F-dwarf planet to require more incoming stellar radiation to raise its global mean surface temperature to a level on par with the G-dwarf planet. In contrast, the M-dwarf planet's atmosphere and surface absorb more radiation, resulting in a smaller instellation required to yield a similar climate to the G-dwarf planet\textemdash only 88\% of the modern solar constant, assuming a 24-hr rotation period for all planets. 

Previous work determined that the instellation value required for global ice cover on an F-dwarf planet was 98\% of the modern solar constant, while an M-dwarf planet was found to require 90\% of the modern solar constant from its star to exhibit a climate similar to modern-day Earth \citep{Shields2013, Shields2014}. The higher instellation value for a frozen F-dwarf planet (107\% instellation, shown in Figure 1a) and lower value for Earth-like conditions on the M-dwarf planet in this work are due to the weighting of the water ice and snow albedos by the spectra of our host stars here, yielding lower near-IR and higher visible ice and snow albedos for M- and F-dwarf spectra, respectively\textemdash rather than using the default albedo parameterization in the GCM. 

A fixed substellar point and lower relative thermal emission on the synchronously-rotating M-dwarf planet keeps temperatures warm on the dayside. However, the planet exhibits a lower maximum dayside surface temperature than the global mean surface temperature of the rapidly-rotating planet. Previous work found that weakened low-latitude zonal winds cool synchronously-rotating planets \citep{Edson2011}, which have been shown to exhibit lower dayside minimum surface temperatures than those on rapidly-rotating planets \citep{Shields2018}. These differences, as well as a greater amount of cloud cover at the substellar point, which has been shown to cool synchronously-rotating planets \citep{Yang2013}, contribute to the resulting energy budget of this planet, with a smaller percentage of its incoming radiation absorbed both in the atmosphere and at the surface compared to the rapidly-rotating case, cooling temperatures. 

We held the atmospheric CO$_2$ concentration fixed in all of our simulations, to isolate the effect of host star SED on a planet's global energy budget. The existence of an active carbon cycle on exoplanets is uncertain, and would certainly affect the energy budgets. If such a cycle operates as it does on the Earth, where the silicate weathering rate is adjusted with temperature (see, e.g., \citealp{Walker1981}), it can be expected that the stronger radiative response to increases in CO$_2$ for M-dwarf planets (see, e.g., \citealp{Shields2013}) may lead to increased fluxes within their absorption budget pathways farther out in their stars' habitable zones compared to G- and F-dwarf planets receiving equivalent instellation. However, the increased instellation required to produce similar climates on G- and F-dwarf planets may help to match any amplified radiative response to increases in CO$_2$ on M-dwarf planets. 
\section{Conclusions} \label{sec:conc}
Using a 3D GCM to calculate energy budget ``Trenberth" diagrams for planets orbiting F-, G-, and M-dwarf stars, we have shown that the spectral energy distribution of a host star heavily influences the energy budget of an orbiting planet. An M-dwarf planet requires 12\% less instellation than a G-dwarf planet to exhibit a climate similar to modern-day Earth, while an F-dwarf planet requires 8\% more instellation, assuming a 24-hr rotation period and fixed CO$_2$. The atmosphere and surface of the M-dwarf planet absorb 12\% more incoming flux than a G-dwarf planet and 17\% more flux than the F-dwarf planet, compensating for the reduced instellation. The spectral dependence of ice and snow albedo, with both absorbing strongly in the near-IR where M dwarfs emit strongly, while heavily reflecting in the visible and near-UV where brighter stars emit, along with CO$_2$ and H$_2$O in the atmosphere absorbing mainly in the near-IR, are responsible for this difference in energy budgets and resulting instellation requirements for planets orbiting different types of stars. For synchronously-rotating M-dwarf planets, smaller flux absorption in the atmosphere and on the surface results in lower dayside minimum/maximum dayside surface temperatures compared to those on M-dwarf planets with 24-hr rotation periods receiving equivalent instellation, with a dayside mean surface temperature that is 37 K colder than its rapidly-rotating counterpart. Should an active carbon cycle exist on exoplanets, the stronger radiative response to increases in CO$_2$ for M-dwarf planets may be matched by the increased instellation required to generate equivalent climates on planets orbiting stars with more visible and near-UV output.
\pagebreak

\section{Acknowledgments} \label{sec:acknowl}

This material is based upon work supported by NASA  under grant number  NNH16ZDA001N, which is part of the ``Habitable Worlds" program, by the National Science Foundation under Award 1753373, and by a Clare Boothe Luce Professorship supported by the Henry Luce Foundation. This research was also supported in part by the National Science Foundation under Grant No. NSF PHY-1748958, and was performed as part of the NASA Astrobiology Institute's Virtual Planetary Laboratory under Cooperative Agreement Number NNA13AA93A. We would like to acknowledge high-performance computing support from Cheyenne (doi:10.5065/D6RX99HX) provided by NCAR's Computational and Information Systems Laboratory, sponsored by the National Science Foundation. We thank the reviewer of this manuscript for providing helpful feedback.
\newpage

\end{document}